\documentclass{iopconfser}
\usepackage{amsmath,amsfonts}
\usepackage{algorithmic}
\usepackage{algorithm}
\usepackage{array}
\usepackage[caption=false,font=normalsize,labelfont=sf,textfont=sf]{subfig}
\usepackage{textcomp}
\usepackage{stfloats}
\usepackage{url}
\usepackage{booktabs}
\usepackage{multirow}
\usepackage{verbatim}
\usepackage{xcolor}
\usepackage{graphicx}
\usepackage{cite}

\begin{document}

\title{KIT Superconducting Undulator Development\\
Story of a successful industrial collaboration \& future prospects}

\author{Bennet Krasch$^{1}$, A. Bernhard$^{1}$, E. Bründermann$^{1}$, S. Fatehi$^{1}$, J. Gethmann$^{1}$, N. Glamann$^{1}$, A. Grau$^{1}$, A. Hobl$^{2}$, A.-S. Müller$^{1}$, D. Saez de Jauregui$^{1}$, E. Tan$^{3}$, and W. Walter$^{2}$}

\affil{$^1$Institute for Beam Physics and Technology (IBPT), Karlsruhe Institute of Technology (KIT), Karlsruhe, Germany}
\affil{$^2$Bilfinger Nuclear \& Energy Transition (Bilfinger) GmbH, Oberhausen, Germany}
\affil{$^3$Australian Synchrotron-ANSTO, Clayton, Australia}

\email{bennet.krasch@kit.edu}

\begin{abstract}
Undulators are X-ray sources widely used in synchrotron storage rings and free-electron laser facilities.
With the commercial availability of low-temperature superconductors, a new type of undulator was born, the superconducting undulator (SCU).
In this context, the industrial cooperation between the Karlsruhe Institute of Technology and Bilfinger Nuclear \& Energy Transition GmbH started more than 15 years ago.
Since then, many projects have been successfully completed, leading to the production of the world’s leading full-scale commercial SCUs based on conduction cooling.
Starting with the SCU15, the first of its kind installed SCU providing light to a beamline, followed by the SCU20 installed and still in operation at the Karlsruhe Research Accelerator.
The successful realisation of such SCUs has required the simultaneous development of appropriate measurement facilities such as CASPER I and CASPER II.
\end{abstract}

\section{Introduction}
Insertion devices (IDs), such as undulators, are crucial for free-electron laser (FEL) facilities and advanced synchrotron radiation sources, as they generate coherent photon beams with high brilliance and with a broad spectral range.
Currently, the most prevalent undulator technology is the permanent magnet undulator (PMU).
However, superconducting undulators (SCUs) have been developed through the commercial application of low-temperature superconducting (LTS) materials.
As the current standard, NbTi wire is wound around coils.
However, Nb$_3$Sn wires are increasingly being used for higher magnetic field generation, where coil handling and winding is more complex and error-prone as special heat treatment is required \cite{PRESTEMON_WIRES}.
It has long been known that theoretically SCUs with the same geometry, such as magnetic gap or period length, have a higher magnetic peak field on the symmetry axis than cryogenic PMUs \cite{CASALBUONI_SCU20, CASALBUONI_SCUdevelopment}.
This fact has also been demonstrated experimentally on several occasions \cite{BAHRDT_VERGLEICH}.
This technology therefore holds great potential not only for current accelerators and beamlines, but also for future storage rings and linear accelerators such as FELs and laser plasma accelerators (LPAs).
A secondary advantage of the SCU is that, unlike cryogenic PMUs, there is no risk of performance degradation due to radiation damage \cite{TISCHER_DAMAGE, SASAKI_DAMAGE, NUHN_DAMAGE, BIZEN_DAMAGE, HUANG_DAMAGE}.
The research and development of SCU technology has been pursued at the Institute for Beam Physics and Technology (IBPT) at the Karlsruhe Institute of Technology (KIT) for over 20 years.
SCUs have been successfully used in continuous operation at the Karlsruhe Research Accelerator (KARA) for 10 years now \cite{CASALBUONI_SCUdevelopment}.
This successful conversion of a research idea into a commercial product was only possible in cooperation with an industrial partner.
The fusion of IBPT research and Bilfinger Nuclear \& Energy Transition GmbH (Bilfinger) industrial know-how  created a coherent partnership with the goal to develop and sell full-scale SCUs without cryogenic refrigerants for current and next generation low-emittance light sources.
Thanks to the resulting synergies the commercial success followed. 

The standard design parameters for each SCU developed by Bilfinger are NbTi wound coils and cryogen-free conduction cooled operation at 4\,K.
This cooling method has the advantage of being user-friendly and particularly attractive for facilities without helium recovery, as only water and electricity are required for operation.
The scientific objective in the development of such SCUs has always been to achieve good field quality in order to maximise the spectral response.
To achieve this, two conditions had to be met.
First, the undulators had to be manufactured commercially with the necessary mechanical precision and accuracy, so Bilfinger developed, improved and perfected new manufacturing processes.
Second, the IBPT has designed, operated and is continuously developing appropriate measurement technologies and adequately equipped measurement facilities to characterise undulators in terms of their performance and magnetic properties.
The results obtained were then fed back into the next iteration at Bilfinger to optimise the ma\-nu\-fac\-tu\-ring process and improve the magnetic characteristics.
In addition, the IBPT offers the unique opportunity to install and test the undulator during the operational phase with KARA .
This provides even a deeper insight into standard operation and therefore allows a much better assessment of the entire SCU technology.

This is the cornerstone of the successful collaboration and technology transfer of innovations from basic research to industry.
As Bilfinger is the worldwide leader in the production of full-scale commercial undulators based on conduction cooling, this development is well documented, see Table \ref{tab:IDs}.
\begin{table}
    \centering
    \caption{\textbf{Portfolio of Bilfinger.} The range of applications for IDs extends from THz to X-rays.}
   \begin{footnotesize} \begin{tabular}{lccccccccc}
    \toprule
      & \multirow{2}{*}{SCU15} & \multirow{2}{*}{SCU20} & HEX70 & ANSTO & S- & \multirow{2}{*}{SCUF} & DLS  & S-PRESSO  & ANSTO  \\
     &  &  & SCW & SCU16 & PRESSO &  & SCW & Mock-Up &  SCW  \\
     \midrule
    period & \multirow{2}{*}{15} & \multirow{2}{*}{20} & \multirow{2}{*}{70} & \multirow{2}{*}{16} & \multirow{2}{*}{18} & \multirow{2}{*}{65} & \multirow{2}{*}{48} & \multirow{2}{*}{18} & \multirow{2}{*}{48}  \\
    length & & & & & & & & \\
    full & \multirow{2}{*}{100.5} & \multirow{2}{*}{74.5} & \multirow{2}{*}{29} & \multirow{2}{*}{98} & \multirow{2}{*}{108*2} & \multirow{2}{*}{20} & \multirow{2}{*}{22.5} & \multirow{2}{*}{14} & \multirow{2}{*}{40}  \\
    periods & & & & & & & & \\
    max. field & \multirow{2}{*}{0.73} & \multirow{2}{*}{1.19} & \multirow{2}{*}{4.3} & \multirow{2}{*}{1.1} & \multirow{2}{*}{1.82} & \multirow{2}{*}{0.88} & \multirow{2}{*}{4.2} & \multirow{2}{*}{2.02} & \multirow{2}{*}{4.6}  \\
    on axis (T) & & & & & & & & \\
    K-value & \multirow{2}{*}{1.0} & \multirow{2}{*}{2.2} & \multirow{2}{*}{28.1} & \multirow{2}{*}{1.6} & \multirow{2}{*}{3.1} & \multirow{2}{*}{5.3} & \multirow{2}{*}{18.8} & \multirow{2}{*}{3.4} & \multirow{2}{*}{20.6}  \\
    (approx.) & & & & & & & & \\
    \multirow{2}{*}{location} & KARA & KARA & NSLS II & AS & EuXFEL & FLUTE & DLS & EuXFEL & AS\\
    & KIT & KIT & BNL, US & AUS & DE & KIT &  UK & DE & AUS\\
    \multirow{1}{*}{status} & delivered & delivered & delivered & delivered & in pro- & in pro-  & in pro- & cold & in pro- \\
     & 2014 & 2017 & 2022 & 2022 & duction & duction & duction & tested  & duction \\
     beam stay & \multirow{2}{*}{7 (15)} & \multirow{2}{*}{7 (15)} & \multirow{2}{*}{8} & \multirow{2}{*}{6} & \multirow{2}{*}{5} & \multirow{2}{*}{35} & \multirow{2}{*}{8} & \multirow{2}{*}{5} & \multirow{2}{*}{6}  \\
     clear (mm) & & & & & & & & \\
     \bottomrule
    \end{tabular}
   \end{footnotesize}
    \label{tab:IDs}
\end{table}
In the following we highlight the decisive aspects and different perceptions of the three partners involved:  the research institute IBPT, which develops SCUs, the company Bilfinger, which manufactures and produces the devices, and the Australian Nuclear Science \& Technology Organisation (ANSTO), which ordered and operates the product as the end customer.

\section{Success due to unique infrastructure at KIT}
The research and development of SCUs began at IBPT more than 20 years ago.
It was clear from the outset that, in addition to industrial cooperation, the further development of this technology could only succeed with the aid of high-precision magnetic measuring systems capable of resolving magnetic field changes in the mT range at 4\,K with µm spatial resolution.
These are the permanently installed CASPER I \cite{MASHKINA_CASPERI} and CASPER II \cite{GRAU_CASPERII} (\textbf{C}h\textbf{a}racterization \textbf{S}etu\textbf{p} for Field \textbf{E}rror \textbf{R}eduction) facilities, continuously improved and in operation for over 10 years, and one new mobile measurement system \cite{GRAU_CASPERII_MOBIL} used for on-site measurements.
\subsection{Magnet and Cryogenics Facilities (MCF)}
\begin{figure*}[t!]
    \centering
    \includegraphics{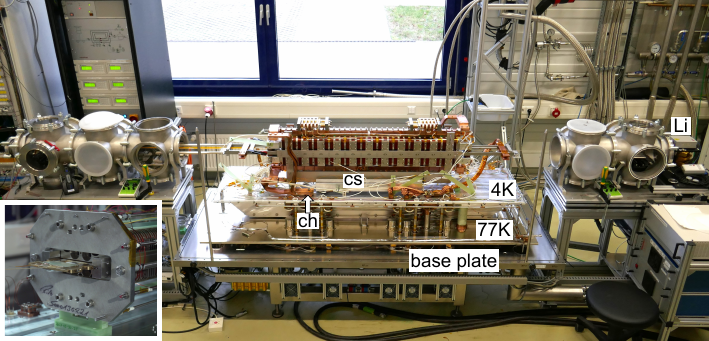}
    \caption{\textbf{CASPER II setup.} Open cryostat with the \textbf{base plate}, followed by a 77\,K cold shield and the 4\,K table to which the coils (HEX70 wiggler) \cite{TANABE_HEX70}) are mounted via a special coil support (\textbf{cs}). Cold heads (\textbf{ch}) with each 1.5\,W at 4\,K. On the left and right are vacuum chambers containing the measurement instrumentation. The laser interferometer (\textbf{Li}) is on the right side. The inset shows the  sledge.}
    \label{fig:CASPERII}
\end{figure*}

At CASPER I \cite{MASHKINA_CASPERI}, short undulator coils up to 500\,mm can be tested and trained in a vertical liquid helium bath cryostat.
The local magnetic field is measured by hall probes mounted on a sledge, which is moved along the beam axis of the undulator via a linear stage.
The local accuracy of 2\,µm is determined by the readout of a linear encoder.
In addition at CASPER II \cite{GRAU_CASPERII}, SC coils up to 2\,m in length can be characterised at 4\,K in a horizontal, cryogen-free, conduction cooled environment by local and integral magnetic field measurement techniques, see also Figure~\ref{fig:CASPERII}.
The local measurements are performed by hall probes on a sledge which is pulled along the coils while guided by two precisely machined rails.
The position of the sledge is measured with sub-µm precision by a laser interferometer.
A moving wire (CuBe, 125\,µm diameter) method \cite{GRAU_TRAINING} provides the first and second field integrals with further optimisation by using attached field correction coils.
A power supply with up to 1500\,A is available at CASPER\,I and II.
Additionally, both systems are equipped with an in-house developed quench detector system from the Institute of Data Processing and Electronics (KIT-IPE). 
The Physical Property Measurement System (PPMS, \url{https://qdusa.com/products/ppms.html}) 9\,T from Quantum Design at the Institute for Technical Physics (KIT-ITEP) enables a direct in-house calibration with an accuracy of 100\,µT for the hall sensors used.
Thus, no further correction factors are necessary.
CASPER I \& II are constantly improved to meet new requirements.
For example, the height of the sledge was reduced to 5\,mm (including the hall sensor) and the width to 104\,mm, see Figure \ref{fig:CASPERII} bottom left corner.
Further reduction to 4\,mm in height (see Figure~\ref{fig:MOBILE_full}) and 65\,mm in width will allow for even smaller beam pipe gaps.
The adjustment of the interferometer at CASPER II by two mirrors was also replaced by a hexapod to make it simpler and more stable.

The variety of performed characterisations of different types of SC magnets \cite{GRAU_FIRSTTESTS, GERSTL_CASPERII, GRAU_30CMCOILS, GRAU_SCU20_2, GRAU_TRAINING, GRAU_SCU20, CASALBUONI_DOUBLING, GRAU_RACETRACK, BAADER_S-PRESSO}, ranging from in-house systems over applications at national and international research institutions, such as CERN, to companies, and the associated publications in scientific journals underline the unique possibilities of these measuring facilities at IBPT.
CASPER I \& II within the MCF form the basis and foundation without which the technology transfer would not have been possible.

\subsection{Mobile measurement systems}
\begin{figure*}[t!]
    \centering
    \includegraphics[width=0.8\textwidth]{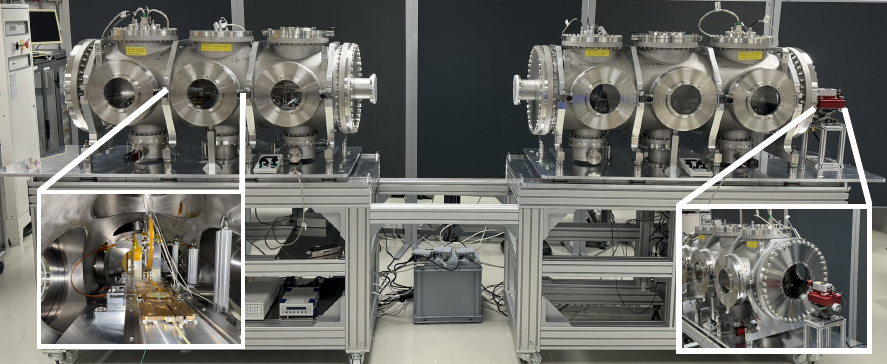}
\caption{\textbf{Newly designed and assembled mobile measurement system.} On the right and left side are the vacuum chambers displayed which are mounted to the final cryostat on site of the end costumer. 
Bottom left-hand corner:  sledge  and (cf.~Fig.~\ref{fig:CASPERII}) the associated instrumentation for pulling, guiding and reading the Hall sensor. Bottom right-hand corner: laser interferometer.}
    \label{fig:MOBILE_full}
\end{figure*}

At IBPT we develop unique, high-sensitive, and also mobile magnetic field measurement setups to precisely evaluate cryogenic IDs in the final cry\-os\-tat.
Measurement techniques for local and integral magnetic field studies are implemented for multipurpose use with any type of magnetic system.
The set ups are characterised by high accuracy and reproducibility, comparable to the stationary CASPER I \& II test stands.
In 2022, the world's first of a kind mobile moving wire system was developed, assembled, tested and referenced at IBPT \cite{GRAU_CASPERII_MOBIL}.
The scientific importance of the system was demonstrated at the Australian Synchrotron (AS) in Clayton for the qualification of the SCU16 \cite{TAN_ANSTO}.
The reproducibility of the 1$^{\mathrm{st}}$ and 2$^\mathrm{nd}$ field integrals was 5x10$^{-6}$ using the moving wire technique (CuBe wire with 125\,µm diameter).
We are now in the assembly phase of the complete upgraded system combining both local and integral magnetic field measurement techniques, see Figure \ref{fig:MOBILE_full}.
This new system will be used for quality assurance (QA) of the new wiggler for the Diamond Light Source during the site acceptance test (SAT) at Bilfinger in Würzburg, which is scheduled for the end of 2024.
In 2025, the system shall be used for the second wiggler ordered by AS.  
In parallel, new measurement techniques such as constant current and pulsed wire methods are tested and evaluated to continuously upgrade and improve the mobile measurement devices.

\subsection{KARA storage ring at KIT}
One crucial aspect is the unique infrastructure in the Accelerator Technology Platform (ATP) at KIT with its various institutes, laboratories and workshops available to us.
The associated expertise in the fields of cryogenic technology, ultra-high vacuum, solid state physics, measurement technology, electronics, but also modelling and simulation is exceptional and includes the aforementioned hall sample calibration at KIT-ITEP and quench detection systems from KIT-IPE.
Furthermore, the KARA storage ring with the associated beamlines made it possible to develop, advance and integrate the SCU technology in an accelerator environment under real-world operations \cite{CASALBUONI_SCU20KARA}. 
This uniquely allows to combine application-oriented ID research, demands for trouble-free accelerator operations, and highest possible brilliance at specific energies in experiments in photon science experiments at beamlines.
For accelerator operations the SCUs should be transparent, so that the electron orbit in a storage ring experiences only negligible disturbances during magnet field tuning.
Preferentially, users should tune the field while at the beamline.
The downtime after a quench should also be very short. 
Besides an excellent magnetic field quality, a good cooling concept, reliable quench safety, high systemic technical requirements for automation and synchronisation are needed.
Our environment at KIT allows us to develop SCUs and to achieve their safe, stable operation, release and control from both, the beamline and the accelerator's control room.
As the complexity of the individual components of a SCU is very high, specialist knowledge from the most diverse areas of research and science is required.
This accumulated knowledge can now not only be directly activated for the continuous further development of SCUs but also opens up a spectrum of completely new research topics, as the following section will show.

\subsection{Additional R\&D and new research areas using synergy}
The IBPT is also driving forward several R\&D projects that are unique in their nature thanks to the existing infrastructure at KIT and in particular at IBPT.
These include in detail worldwide exclusive R\&D on compact, sustainable, energy- and resource-saving SC magnets and SCUs, based on high-temperature (HTS) technology.
One promising approach is the development of SCUs by stacking 30 non-insulated, laser-structured, 12\,mm HTS tapes to realise smaller period lengths $\leq$ 8\,mm \cite{PRESTEMON_TAPE, HOLUBEK_TAPE, WILL_TAPE, KRASCH_MT}.
Another approach is a helical undulator with 4\,mm non-insulated HTS tapes for compact FELs \cite{RICHTER_HELICAL}.
First cooling tests in liquid nitrogen were carried out to show the success of both methods.
By using non-insulated, so called not stabilised, HTS tapes, can have the advantage that they can be operated quench-free, i.e. well above $I_\mathrm{c}$.
In addition, the realisation of thermal transitions for cryogenic vacuum chambers \cite{CHA_THERMAL}, SC magnets for a short-length electron transport line for LPAs \cite{FATEHI_MAGNET1, FATEHI_MAGNET2, BERNHARD_MAGNETS} or the R\&D on THz undulators for the European XFEL \cite{GETHMANN_THZ} and for FLUTE (Far Infrared Linac and Test Experiment) \cite{GETHMANN_FLUTE} are carried out.
Since the emitted synchrotron radiation for protons of the Future Circular Collider (FCC-hh) is similar to electrons at KARA, CERN selected the KIT accelerator test facility. Several prototypes were manufactured and installed in the beam screen test-bench experiment at KARA \cite{GONZALEZ_BESTEX}.
The development of SCUs with the possibility to change the period length during operations is also studied \cite{HOLUBEK_SWITCH, CASALBUONI_SWITCH}.
The responsible and efficient use of energy and materials is an important topic for the current and future R\&D on SCUs and their sustainable operation.
The new HTS SCU concepts have the potential for reducing the energy consumption for cooling.
The KIT Test field for Energy Efficiency and Grid Stability in Large-Scale Research Infrastructures (KITTEN) is a unique union of two research infrastructures, KARA and the Energy Lab 2.0, which supports research from the component to the system level \cite{GETHMANN_KITTEN}.
This year, in 2024, a KIT-led project, the Research Facility 2.0, has also started to address the sustainability of particle accelerator \cite{RF2Website}.

\section{Bilfinger’s full-scale conduction cooled SC undulators portfolio}
\begin{figure}[h]
    \centering
    \includegraphics{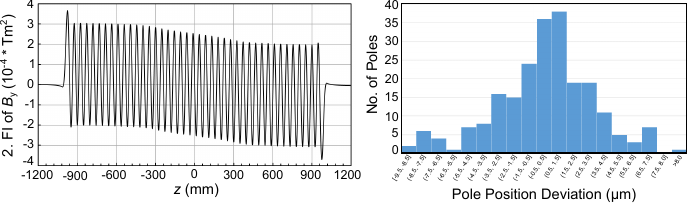}
    \caption{\textbf{(a) Computed $2^\mathrm{nd}$ field integral for a 40 full-period wiggler. (b) Distribution of the pole position deviation from the target position of a 2\,m long undulator coil.}}
    \label{fig:BNET_FT_Pole}
\end{figure}

All IDs need to be transparent for the electron beam of the accelerator they are operating in.
Therefore, stringent requirements are usually set for the 1$^{\mathrm{st}}$ and 2$^\mathrm{nd}$ field integrals concerning all operating currents.
The magnetic design of all Bilfinger’s recent IDs is point-symmetric, with a ¼ - ¾ end field adaption to achieve a straight trajectory around the target orbit inside the device.
The design approach iterates the number of turns in the last three winding grooves of the ID's coils.
The last step of optimisation of the trajectory, both in design and in operation, is achieved with so-called auxiliary coils in the last grooves.
These auxiliary coils are wound with a small, typically 0.25\,mm, SC wire to maintain the operating currents within a few amperes.
The main coils’ multifilamentary NbTi SC wire is of 0.68\,mm\,x\,1.08\,mm rectangular cross section, therefore operating currents range from 300\,A to above 1000\,A.
A typical design result for the $2^\mathrm{nd}$ field integral is shown in Figure \ref{fig:BNET_FT_Pole} (a).
In the course of the winding former manufacturing optimisation, improved methods have been found to mechanically machine the winding formers from one piece each, even if slightly longer than 2\,m.
The positions of the winding former’s iron poles are crucial to achieve low field and phase errors.
With the present manufacturing method, the deviations from the target positions are typically within $\pm$\,10\,µm, as measured by a coordinate-measuring machine.
An example of the position accuracy achieved within a 222 pole winding former is shown in Figure \ref{fig:BNET_FT_Pole} (b).
After coil winding, the SC wires have to be fixed to withstand the high Lorentz force densities during operation, therefore, the coils are vacuum-pressure impregnated.
The process optimised during the last years is profiting from a resin-mixing plant that allows preparing two-component resin mixtures from independently pre-heated and evacuated storage vessels on demand, reducing the risk of premature gelling of the resin during the mould filling process.

Experience from the SCU16 BioSAXS (Biological Small Angle X-Ray Scattering) beamline \cite{TAN_BRIGHT} and the need for high mechanical accuracies in the EuXFEL project S-PRESSO \cite{CASALBUONI_SCU} gave the impact to re-design the mechanical structure around the SC coils (see also Table \ref{tab:IDs}).
Stiff plates provide the moment of inertia to counter-act bending of the $\sim$4.5\,m long cold mass of S-PRESSO.
As a side effect, the new mechanical structure also allows a more accurate positioning of the cold electron beam chamber inside the gap of the magnets.
This is of particular importance, whenever the beam chamber has to act as the guiding element for cold local magnetic measurements inside the ID.

Thorough design, manufacturing and assembly of all cryogenic components of an ID is essential for reliable and efficient operation.
Based on the experience of IDs being or having been operational at KIT, BNL, and ANSTO, the cryogenic design has been improved towards a more standardised and modular approach.
Identical cryocooler, thermal bus, and current lead sub-assemblies are now foreseen for all IDs presently under design or manufacture. 
The design uses finite element software packages like ANSYS and nonlinear, isotropic or orthotropic materials’ data for the simulations.
An example of electron beam tube temperatures during operation with heat loads from thermal conduction and beam induced losses is shown in Figure \ref{fig:AS_SCU16} (a).
The temperature of the beam tube inside the 4\,K magnet is typically below 20\,K.
This modular cryogenic design and assembly approach helps reducing operational risks and time for assembly.
As a result of this cryogenic concept, the cooldown of these SC systems is just a start-button-operation, no handling of cryogens is required, and there is no need for any cryogen-relief safety elements.

\section{Experience of an end customer: SCU16 at the Australian Synchrotron (AS)}
\begin{figure}[h]
    \centering
    \includegraphics{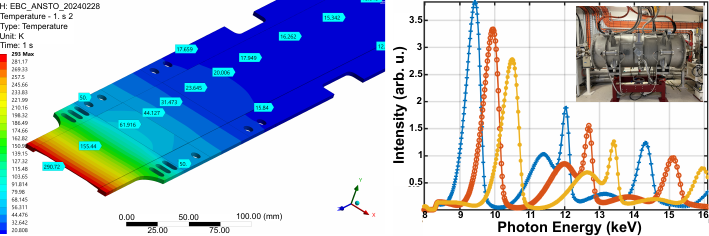}
    \caption{\textbf{SCU16.} (a) Temperature profile of electron beam chamber during operation as computed with ANSYS. (b) Spectrum at 0.95\,T (blue), 1.00\,T (red), and 1.05\,T (yellow) (inset: installed at AS).}
    \label{fig:AS_SCU16}
\end{figure}

The AS \cite{LEBLANC_AS} is a 3\,GeV user facility operated by ANSTO since 2005.
Based on the successful long-term operation of SCU20 at KARA, the BioSAXS beamline selected an SCU with a photon energy of 12.4\,keV at the fifth harmonic to be designed and built by the company Bilfinger \cite{TAN_ANSTO}.
The SCU16 is a vertical racetrack undulator with a period length of 16\,mm, a magnet length of 1.6\,m and the longest SCU to be installed in a light source (cf.~Table \ref{tab:IDs}). 
The maximum B-field on axis is 1.084\,T at $K=1.62$ with a magnetic gap of 8.0\,mm and a vacuum chamber height of 6\,mm. 
First field measurements with the CASPER II system were performed at KIT prior to installation.
In 2022, the SCU16 was installed and commissioned at AS.
It took less than 4 days to cool down the entire system with an insulation vacuum of $10^{-6}$\,mbar.
In addition, only up to 10 quench cycles were required to restore the magnetic field after installation in the storage ring.
The maximum ramp rate for the main current is 4.5\,A/s, which means that the maximum field can be reached within 3\,min.
In the case of a quench, the temperature at the magnets/diode sensors rises to 20\,K and it takes 25\,min to reach the operational temperature again. 
With the mobile measurement system developed at KIT, the field quality was checked on site using the moving wire method and it was shown that the transport had no influence on the field quality.
There is a maximum orbit deviation of less than 100\,µm with a maximum disturbance of less than 5\,µm with orbit feedback enabled.
The overall change in horizontal and vertical tunes was $v_{x, y}$= +0.0010, +0.0018 and is similar to other in vacuum undulators at the AS.
The spectrum of the SCU is displayed in Figure \ref{fig:AS_SCU16} (b).
It can be summarised that the cooling system operates well below the design temperatures, providing more than adequate performance for the BioSAXS beamline ($>10^{12}$\,ph/s on sample) with quench recovery time of under 30\,min.

\section{Conclusion \& outlook}
Despite SCUs having a shorter development history than CMPUs, the strong collaboration between KIT and its industrial partner Bilfinger has advanced their development to a successful commercial product.
Today, Bilfinger is the world market leader in the production of full-scale commercial SCUs worldwide.
This was possible because we have continuously worked on improving and optimising magnetic cryogenic characterisation facilities at KIT-IBPT to enable precise and accurate field measurements.
These systems are also unique worldwide.
Furthermore, Bilfinger has continuously optimised its manufacturing processes to ensure the desired precision of the SCUs.
Ultimately, the end customers and photon science experiments benefit from this technology transfer.
Building on the knowledge gained, the IBPT is not only further developing advanced and unique measurement technologies for current and future IDs used in X-ray light sources, but also driving forward developments of SC THz undulators and HTS systems including SCUs, magnets for compact transport and injection lines, and many more.

\end{document}